# Spin injection into vanadium dioxide films from a typical ferromagnetic metal, across the metal-insulator transition of the vanadium dioxide films

Kazuma Tamura[1,a], Teruo Kanki[2], Shun Shirai[1], Hidekazu Tanaka[2], Yoshio Teki[3], and Eiji Shikoh[1,a]

[1]*Graduate School of Engineering, Osaka City University, Osaka 558-8585, Japan*
[2]*Institute of Scientific and Industrial Research, Osaka University, Ibaraki, Osaka 567-0047, Japan*
[3]*Graduate School of Science, Osaka City University, Osaka 558-8585, Japan*

[a]Corresponding authors: tamura@mc.elec.eng.osaka-cu.ac.jp (K.T.), shikoh@eng.osaka-cu.ac.jp (E.S.)

A vanadium dioxide $VO_2$ film shows metal-insulator transition (MIT) induced by changing environmental temperature. We report the temperature dependence of electromotive force properties generated in $VO_2$/$Ni_{80}Fe_{20}$ bilayer junctions under the ferromagnetic resonance (FMR) of the $Ni_{80}Fe_{20}$ layer. An electromotive force generated in a $VO_2$/$Ni_{80}Fe_{20}$ bilayer junction under the FMR showed a small change across the MIT temperature of the $VO_2$ film, while the $VO_2$ film resistance drastically changed. This behavior was not only explained with the temperature dependence of the electromotive force property generated in the $Ni_{80}Fe_{20}$ film itself under the FMR, but also with the generated electromotive forces due to the inverse spin-Hall effect (ISHE) in the $VO_2$ film under the FMR of the $Ni_{80}Fe_{20}$ film. That is, we successfully demonstrated the spin injection from a $Ni_{80}Fe_{20}$ film into a $VO_2$ film across the MIT temperature of the $VO_2$ film.

To switch a pure spin current, which is a flow of spin angular momenta and a dissipation-less information propagation, by using external ways such as applying a voltage, light irradiation or heating, is an indispensable issue for future spintronic application, like spin-transistors.[1-3] Up to now, a Si-based spin transistor operated by applying a voltage has been demonstrated,[3] and to switch a spin current through light irradiation have been challenged. Meanwhile, there are no studies for switching a spin current with an environmental temperature change, not due to the so-called spin-Seebeck effect.[4] In this study, to switch a spin current by a temperature change, a vanadium dioxide $VO_2$ film which shows metal-insulator transition (MIT) with the crystal structure change induced by changing environmental temperature is focused.[5] The MIT in $VO_2$ films usually occurs in the temperature range between 280 K and 350 K, depending on the film thickness,[5,6] and a thinner $VO_2$ film shows an MIT at lower temperature. Recently, the spin injection into a $VO_2$ film from a ferrimagnetic insulator $Y_3Fe_5O_{12}$ (YIG) film by using the spin-pumping was achieved across the MIT temperature.[7] However, for the practical use, to prepare high quality YIG films is needed and basically hard. Thus, in this study, the spin injection from a typical ferromagnetic metal $Ni_{80}Fe_{20}$ film into a $VO_2$ film and the controllability of the spin current in $VO_2$ with environmental temperature change across the MIT are tried. We report the temperature dependence of electromotive force properties generated in $VO_2$/$Ni_{80}Fe_{20}$ bilayer junctions under the ferromagnetic resonance (FMR) of the $Ni_{80}Fe_{20}$ layer.

Our sample structure and experimental set up are illustrated in Figures 1(a) and (b). A spin injection effect into a $VO_2$ film is observed as follows: in $VO_2$/$Ni_{80}Fe_{20}$ bilayer junctions, a pure spin-current, $J_S$, is generated in the $VO_2$ layer by the spin-pumping of the $Ni_{80}Fe_{20}$ induced by the FMR.[8] The generated spin current is converted to a charge current, $J_C$, with the inverse spin-Hall effect (ISHE)[8] in the $VO_2$, which is expressed as,[10]

$$J_C \propto \theta_{SHE}\, J_S \times \sigma, \qquad (1)$$

where $\theta_{SHE}$ and $\sigma$ are the spin-Hall angle in $VO_2$ films and the spin-polarization vector of the $J_S$, respectively. The converted $J_C$ is detected as an electromotive force, $E$, via the sample resistance, $R$. Therefore, the $E$ is expressed as,[9,11]

$$E = R\, J_C \propto R\, \theta_{SHE} J_S \times \sigma. \qquad (2)$$

That is, if the electromotive force due to the ISHE in $VO_2$ is detected under the FMR of the $Ni_{80}Fe_{20}$ layer, it is clear evidence of the achievement of the spin injection from a $Ni_{80}Fe_{20}$ layer into a $VO_2$ layer. We analyze the origins of the obtained electromotive forces under the FMR with some control experiments, and conclude that the spin injection from a $Ni_{80}Fe_{20}$ film into a $VO_2$ film has been achieved and the spin current generation efficiency in $VO_2$ films by the spin pumping has been changed by the MIT.

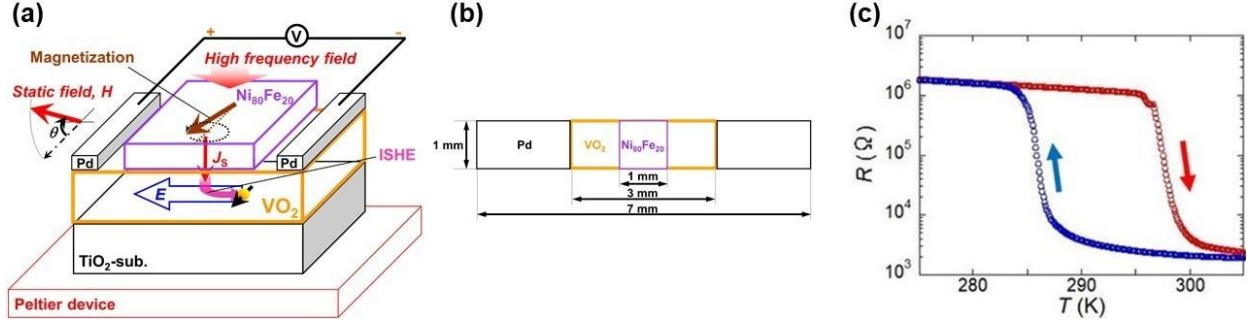

FIG. 1. (a) Bird's-eye-view and (b) top-view illustrations of our sample and orientations of external applied magnetic field $H$ used in the experiments. $J_S$ and $E$ correspond to the spin current generated in a $VO_2$ film by the spin-pumping and the electromotive forces due to the ISHE in a $VO_2$ film, respectively. (c) Temperature dependence of a $VO_2$ film resistance.

$VO_2$ thin films were formed on $TiO_2$(001) single crystal substrates using a pulsed laser deposition technique (ArF excimer: $\lambda$ = 193 nm). The target employed was a sintered vanadium oxide pellet of $V_2O_5$ (Kojundo Chemical Lab. Co., Ltd., 99.9%). The fabrication condition of $VO_2$ films was at a substrate temperature of 450 °C in an $O_2$ gas pressure of 1.0 Pa. This fabrication process is almost same as the previous result where epitaxial $VO_2$ films have been obtained.[6] The $VO_2$ film thickness was set to be 10 nm. After the $VO_2$ depositions, the sample substrate was transferred to another vacuum chamber for the preparation of metal films. This transfer process and the next $Ni_{80}Fe_{20}$ deposition process were implemented as soon as possible to keep the surface state of the $VO_2$ films because of the surface sensitivity to the air. An electron beam (EB) deposition technique was used to deposit $Ni_{80}Fe_{20}$ (Kojundo Chemical Lab. Co., Ltd., 99.99%) to a thickness of 25 nm on the $VO_2$ films with a shadow mask, under a vacuum pressure of <$10^{-6}$ Pa. The deposition rate and the substrate temperature during $Ni_{80}Fe_{20}$ depositions were set to be 0.1 nm/s and an ambient temperature, respectively. Finally, palladium (Pd: Furuuchi Chemical Co., Ltd., 99.99% purity) as electrodes was deposited by EB deposition through another shadow mask, under a vacuum pressure of <$10^{-6}$ Pa. The deposition rate and the substrate temperature during Pd depositions were also set to be 0.1 nm/s and an ambient temperature, respectively.

Electrical properties were evaluated using a two-probe method with a source measure unit (Keithley Instruments, 2614B). The electrical property evaluation was implemented in a small vacuum chamber equipped with a temperature variable stage using a Peltier device. To excite FMR in $Ni_{80}Fe_{20}$ for investigation of the spin injection effect into $VO_2$ films, a coplanar waveguide connected to a vector network analyzer (VNA: KEYSIGHT Technology, N5221A) and a couple of electromagnets were used. A nano-voltmeter (Keithley Instruments, 2182A) to detect electromotive forces from the samples was used. This spin-pumping experiment system also equips with another Peltier device under the sample mount position. Leading wires for detecting the output voltage properties were directly attached at the two Pd electrodes of samples with silver paste.

Figure 1(c) shows a temperature dependence of resistance of a $VO_2$ film. The sample temperature was changed with a rate of 2 K/min, monitoring with a thermocouple. The $VO_2$ film resistance changes by three orders of magnitude and shows a hysteresis in this temperature range, reflecting the typical MIT of $VO_2$ films.[5,6] Moreover, the $VO_2$ resistance is steeply changed which is a typical evidence that the $VO_2$ film is almost epitaxially grown.[6]

The inset in Figure 2(a) shows an FMR spectrum for a $VO_2$/$Ni_{80}Fe_{20}$ bilayer junction at 300 K, where an external magnetic field orientation angle to the sample film plane, $\theta$, is 0° and the frequency of the high frequency electrical current, $f$, is 5 GHz. The FMR field of the $Ni_{80}Fe_{20}$ film, $H_{FMR}$, is 325 Oe, and the saturation magnetization of the $Ni_{80}Fe_{20}$ film, $M_S$, is estimated to be 671 G under the FMR conditions in the case of the in-plane field:[12]

$$\frac{\omega}{\gamma} = \sqrt{H_{FMR}(H_{FMR} + 4\pi M_S)}, \tag{3}$$

where $\omega$ and $\gamma$ are the angular frequency $2\pi f$ and the gyromagnetic ratio of $1.86 \times 10^7$ G$^{-1}$s$^{-1}$ of $Ni_{80}Fe_{20}$, respectively. The main panel of Fig. 2(a) shows output voltage properties of the same $VO_2/Ni_{80}Fe_{20}$ bilayer junction under the FMR of the $Ni_{80}Fe_{20}$ layer at 300 K. Clear output voltage properties have been observed and the voltage sign has been inverted at the magnetization reversal of the $Ni_{80}Fe_{20}$ layer. This voltage sign inversion associated with the magnetization reversal in $Ni_{80}Fe_{20}$ is a characteristic of the typical ISHE.[9,13,14] The solid lines in Fig. 2(a) are the curve fits obtained using the following equation:[9,14]

$$V(H) = V_{Sym} \frac{\Gamma^2}{(H - H_{FMR})^2 + \Gamma^2} + V_{Asym} \frac{-2\Gamma(H - H_{FMR})}{(H - H_{FMR})^2 + \Gamma^2}, \qquad (4)$$

where $H$ is an external static magnetic field and $\Gamma$ denotes the damping constant (18.0 Oe in this study). The first and second terms in the eq. (4) correspond to the symmetry term to $H$ (e.g. the ISHE) and the asymmetry term to $H$ (e.g. the anomalous Hall effect and other effects showing the same asymmetric voltage behavior relative to the $H$, like parasitic capacitances), respectively. $V_{Sym}$ and $V_{Asym}$ correspond to the coefficients of the first and second terms in eq. (4), respectively. On the other hand, Fig. 2(b) shows the output voltage properties of a $Cu/Ni_{80}Fe_{20}$ bilayer junction under the FMR of the $Ni_{80}Fe_{20}$ layer at 300 K. No clear output voltages were observed at the $H_{FMR}$ of the $Ni_{80}Fe_{20}$ layer. Since Cu has small spin orbit interaction to generate the ISHE, this behavior is reasonable. This suggests the observed electromotive forces in the $VO_2/Ni_{80}Fe_{20}$ bilayer junctions under the FMR of the $Ni_{80}Fe_{20}$ layer are mainly due to the ISHE in $VO_2$ films. Also, we can say that the electromotive force generated in the $Ni_{80}Fe_{20}$ layer itself under the FMR is shunted in the Cu layer. This similar shunting effect for the samples with $VO_2$ can also be taken into account to explain the results.

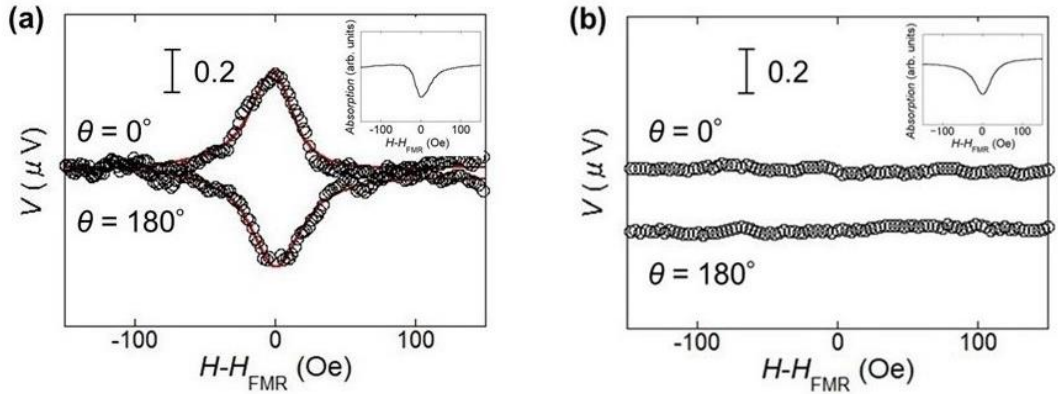

FIG. 2. (a) Output voltage properties of a $VO_2/Ni_{80}Fe_{20}$ bilayer junction under the FMR of the $Ni_{80}Fe_{20}$ layer at 300 K. The inset shows the FMR spectrum for the bilayer junction at 300 K, and at the $\theta$ of 0°. (b) Output voltage properties of a $Cu/Ni_{80}Fe_{20}$ bilayer junction under the FMR of the $Ni_{80}Fe_{20}$ layer at 300 K. The inset shows the FMR spectrum for the bilayer junction at 300 K, and at the $\theta$ of 0°.

Figure 3 shows temperature dependences of the $V_{Sym}$ of a $VO_2/Ni_{80}Fe_{20}$ bilayer sample (circles) and the sample resistance (a solid line). As the $VO_2$ film changed from an insulator to a metal, the $V_{Sym}$ became slightly small. This behavior is not only be explained with the electromotive force property generated in the $Ni_{80}Fe_{20}$ film itself under the FMR[15] since the resistance of $VO_2$ films is drastically changed across the MIT and the ferromagnetic transition temperature is much higher than the temperature range in this study. Thus, to explain the small change of the $V_{Sym}$ against the temperature, the $\theta_{SHE}$ and the injected $J_S$ into $VO_2$ films must be changed by controlling the environmental temperature under the relationship expressed as the eq. (2). This is reasonable because the crystal structure of $VO_2$ is changed across the MIT, that is, the electronic properties in $VO_2$ films are drastically changed across the MIT. The electromotive force generated in the interface between the $VO_2$ and $Ni_{80}Fe_{20}$ films in the $VO_2/Ni_{80}Fe_{20}$ junctions may be considered by using, for example, other MIT materials[5] and it is an interesting study, whereas this is also excluded by the shunting effect in the $VO_2$ film. That is, the ISHE in $VO_2$ films is dominant as origins of observed electromotive forces in the $VO_2/Ni_{80}Fe_{20}$ junctions under the FMR. Thus, we concluded that the spin injection into $VO_2$ films was achieved by the spin-pumping using a $Ni_{80}Fe_{20}$ film.

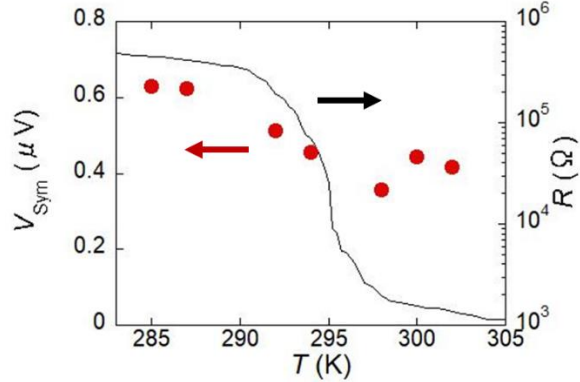

FIG. 3. Temperature dependences of the electromotive forces generated in a VO$_2$/Ni$_{80}$Fe$_{20}$ bilayer junction under the FMR of the Ni$_{80}$Fe$_{20}$ film (circles) and the sample resistance (a solid line).

Finally, in order to investigate the spin current generation efficiency by the spin pumping, we focused on the real part of the spin mixing conductance, $g_r^{\uparrow\downarrow}$, which represents the parameter to determine the spin pumping efficiency at the interface between the Ni$_{80}$Fe$_{20}$ and the VO$_2$ layers. The $g_r^{\uparrow\downarrow}$ is expressed as,[16]

$$g_r^{\uparrow\downarrow} = \frac{4\pi M_S d_F}{g\mu_B}(\alpha - \alpha_0), \quad (5)$$

where $g$, $\mu_B$, $d_F$, $\alpha$, and $\alpha_0$ are the Landé g-factor, the Bohr magneton, the thickness of the Ni$_{80}$Fe$_{20}$ layer, the Gilbert damping constant, and the intrinsic Gilbert damping constant of the Ni$_{80}$Fe$_{20}$, respectively. The $g_r^{\uparrow\downarrow}$ depends on the $\alpha$. In other words, the spin current generation efficiency depends on the $\alpha$. Figure 4(a) shows a frequency dependence of the half width at half maximum resonance linewidth, $\Delta H$, for a VO$_2$/Ni$_{80}$Fe$_{20}$ bilayer junction in the temperature range across the MIT. The $\alpha$ is obtained using the following equation:[17]

$$\Delta H = \Delta H_0 + \frac{2\pi\alpha}{|\gamma|}f, \quad (6)$$

where $\Delta H_0$ is the frequency independent $\Delta H$. Fig. 4(b) shows temperature dependence of the $\alpha$ of a VO$_2$/Ni$_{80}$Fe$_{20}$ bilayer junction (red circles) and a Ni$_{80}$Fe$_{20}$ single layer sample (black circles). The $\alpha$ of a VO$_2$/Ni$_{80}$Fe$_{20}$ bilayer junction increased at a higher temperature, while the $\alpha$ of a Ni$_{80}$Fe$_{20}$ single layer sample is almost constant in this temperature range. Since the $g_r^{\uparrow\downarrow}$ depends on the $\alpha$, we have concluded that the spin current generation efficiency by the spin pumping is affected by the MIT. The temperature dependence of the $\alpha$ of the VO$_2$/Ni$_{80}$Fe$_{20}$ bilayer junction is also an evidence of the spin injection effect from the Ni$_{80}$Fe$_{20}$ layer into the VO$_2$ layer.

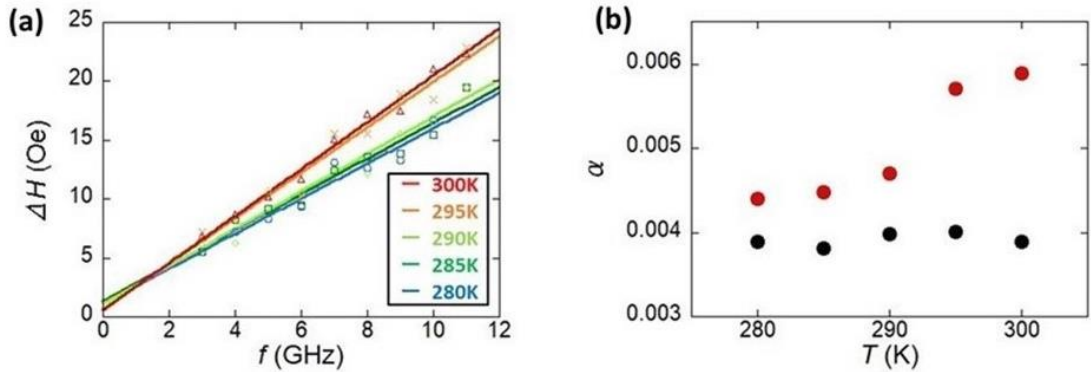

FIG. 4. (a) Frequency dependence of the half width at half maximum resonance linewidth, $\Delta H$, for a VO$_2$/Ni$_{80}$Fe$_{20}$ bilayer junction in the temperature range of the MIT. (b) Temperature dependences of the Gilbert damping constant, α, of a VO$_2$/Ni$_{80}$Fe$_{20}$ bilayer junction and a Ni$_{80}$Fe$_{20}$ single layer sample.

In summary, we reported the temperature dependence of electromotive force properties generated in $VO_2/Ni_{80}Fe_{20}$ bilayer junctions under the FMR of the $Ni_{80}Fe_{20}$ layer. The detected electromotive forces in the bilayer junctions under the FMR showed a small change in the temperature range, while the resistance $VO_2$ films was drastically changed across the MIT. This behavior was not only explained with the temperature dependence of the electromotive force property generated in the $Ni_{80}Fe_{20}$ film itself under the FMR. The ISHE in $VO_2$ films was the dominant origin of the generated electromotive forces in the $VO_2/Ni_{80}Fe_{20}$ bilayer junctions under the FMR, that is, we achieved the spin injection from $Ni_{80}Fe_{20}$ into a MIT compound, $VO_2$. Additionally, the Gilbert damping constant increased at higher temperatures, that is, the spin current generation efficiency by the spin pumping from a $Ni_{80}Fe_{20}$ film into a $VO_2$ film was changed by the MIT.

This research was partly supported by a Grant-in-Aid from the Japan Society for the Promotion of Science (JSPS) for Scientific Research (B) (26286039 (to E. S. and T. K.)), by the Cooperative Research Program of "NJRC Mater. & Dev.," (to E. S., T. K. and H. T.) and by the OCU Strategic Research Grant 2018 for top priority researches (to E. S.).

The data that support the findings of this study are available from the corresponding author upon reasonable request.